\documentclass[aps,pre,reprint, amsmath, amssymb,superscriptaddress]{revtex4-1}

\usepackage{bm}
\usepackage[retainorgcmds]{IEEEtrantools}
\usepackage{graphicx}
\usepackage{mathrsfs}
\usepackage{amsmath}
\usepackage{amssymb}
\usepackage{color}
\usepackage{amsfonts}
\usepackage{nicefrac}

\newcommand{\ud}{\,\mathrm{d}}

\DeclareMathOperator{\tr}{tr}

\newcommand{\JN}{J_{\mathcal{N}}}
\newcommand{\JE}{J_{E}}
\newcommand{\JQ}{J_{Q}}
\newcommand{\FN}{\mathcal{F}_{\mathcal{N}}}
\newcommand{\FE}{\mathcal{F}_{E}}

\begin{document}

\title{Microscopic theory of a non-equilibrium open bosonic chain}
\date{\today}
\author{Jader P. Santos}
\affiliation{Universidade Federal do ABC,  09210-580 Santo Andr\'e, Brazil}
\author{Gabriel T. Landi}
\email{gtlandi@if.usp.br}
\affiliation{Instituto de F\'isica da Universidade de S\~ao Paulo,  05314-970 S\~ao Paulo, Brazil}

\begin{abstract}

Quantum master equations form an important tool in the description of transport problems in open quantum systems. 
However, they suffer from the difficulty that the shape of the Lindblad dissipator depends sensibly on the system Hamiltonian.
Consequently, most of the work done in this field has focused on phenomenological dissipators which act locally on different parts of the system. 
In this paper we show how to construct Lindblad dissipators for quantum many-body systems starting from a microscopic theory of the system-bath interaction. 
We consider specifically a one-dimensional bosonic tight-binding chain connected to two baths at the first and last site, kept at different temperatures and chemical potentials. 
We then shown that, even though the bath coupling is local, the effective Lindblad dissipator stemming from this interaction is inherently non-local, affecting all normal modes of the system. 
We then use this formalism to study the current of particles and energy through the system and find that they have the structure of Landauer's formula, with the bath spectral density playing the role of the transfer integral. 
Finally, we consider infinitesimal temperature and chemical potential gradients and show that the currents satisfy Onsager's reciprocal relations, which is a consequence of the fact that the microscopic quantum dynamics obeys detailed balance.

\end{abstract}
\maketitle{}

%
%
%
%
\section{\label{sec:int}Introduction}
 
Transport phenomena in quantum systems constitutes one of the major areas of research in condensed matter. 
And despite the great experimental and theoretical progress that it has seen in recent years, the number of open questions remains enormous. 
Part of this difficult lies with the fact that non-equilibrium processes generally lack a unified framework. 
Most of the advances in this area have relied on linear response theories such as the Kubo formula \cite{Kubo1957,*Kubo1957a,Mahan,Jeon1995,*Jeon1995a} or the Landauer-B\"utiker formalism \cite{Landauer1987,VanWees1988,Pastawski1991,Baringhaus2014,Datta1997}, which relate non-equilibrium quantities to equilibrium fluctuations.  
However, these formalisms do not allow one to describe the dependence on the structure of the reservoirs, which is essential for systems beyond linear response. 
Moreover, they are not well suited to model specific details of the interaction between the system and the bath. 
For micro and mesoscopic systems, new evidence suggests \cite{Groeblacher2013} that this coupling may be much more complicated than expected. 

An alternative approach to non-equilibrium processes is that of Lindblad quantum master equations \cite{Lindblad1976,Gardiner2004,Breuer2007}. 
When derived from a microscopic model of the system-reservoir interactions, the dynamics generated by this approach satisfies detailed balance and will, when all reservoirs are in equilibrium with each other, take the system to the correct Gibbs thermal state. 
However, the functional structure of these master equations depend sensibly on the form of the Hamiltonian, making them difficult to implement. 
These difficulties have led researchers to focus on phenomenological dissipators which act locally on specific parts of the system \cite{Manzano2012,*Asadian2013,Benenti2009,Karevski2009,*Platini2010,*Platini2008,Karevski2013,*Popkov2013a,Prosen2008,*Prosen2014,Prosen2015,*Prosen2011,*Prosen2011b,Prosen2009,*Prosen2012,*Prosen2013a,popkov1,*Popkov2013b,*popkov2,Mendoza-Arenas2013,*Mendoza-Arenas2014a,*Mendoza-Arenas2013a,Vorberg2015,Yan2009,*Zhang2009,Znidari2014,*Znidaric2015,*Znidaric2011,*Znidaric2013,Landi2014b,Landi2015a}.
The equations generated by this approach are much easier to handle, but will only give physically reasonable results in the limit where the different parts of the system are weakly coupled. 

This was tested in a specific example in Ref.~\cite{Rivas2010}, where the authors compared the  dynamics of two coupled oscillators interacting with local Lindblad dissipators, with exact numerical solutions. They found that the use of local dissipators was justified only when the interaction between the two oscillators is weak. 
This argument was then used recently in Ref.~\cite{Asadian2013} to study the transport properties of a bosonic tight-binding chain connected to two local dissipators acting on the first and last sites. The authors were, among other things, able to compute the particle and energy currents exactly for chains of arbitrary sizes and showed that these currents are ballistic, as expected from harmonic systems. 
However, since they used local dissipators, their results are only valid in the limit where the chain is weakly coupled. 

The goal of this paper is to show how to go beyond the use of these phenomenological local dissipators and derive microscopic Lindblad  dissipators  for  quantum many-body systems.
For concreteness, and also for the purpose of comparison, we consider here also the the bosonic tight-binding chain, with the first and last sites  coupled to two baths  kept at different temperatures and chemical potentials (which we model, as usual, as an infinte collection of bosonic modes). 
We show that even though the coupling is local (the baths couple only to the first and last sites), the Lindblad dynamics generated by this process is inherently  non-local, affecting all normal modes of the system. 
We then study the flow of particles and heat through the system and show that both have the general structure of Landauer's formula, with a transfer matrix depending on the bath spectral density. 
This therefore establishes a connection between the Lindblad and the Landauer--B\"utiker formalism.
Moreover, we find that  in the limit of infinitesimal temperature and chemical potential gradients,
the fluxes  obey  Onsager's reciprocal relations, which is in agreement with the fact that the dynamics satisfies detailed balance.

The approach that will be used here is quite general and extends far beyond the tight-binding model. 
In principle, it is applicable to any system amenable to a quasi-particle diagonalization (in the language of second quantization) and which is coupled linearly (in the creation and annihilation operators) to an arbitrary number of heat baths. 
It also encompasses interacting systems described using mean-field theory. 
The present work constitutes a first step towards a more systematic method of constructing Lindblad dissipators for quantum many-body systems.

%
%
%
%
\section{\label{sec:framework}Formal Framework}

We consider a 1D lattice with $N$ sites, each described by a bosonic operator $a_n$, where $n = 1,\ldots, N$. 
The system is assumed to evolve according to the tight-binding Hamiltonian 
\begin{equation}\label{H1}
H = \epsilon \sum\limits_{n=1}^N a_n^\dagger a_n - \frac{g}{2} \sum\limits_{n=1}^{N-1} (a_{n}^\dagger a_{n+1} + a_{n+1}^\dagger a_n)
\end{equation}
The first term is the on-site energy and the second term is the hopping (tight-binding) term describing the tunneling of bosons through different sites. 
We also assume that the system is coupled to two heat baths connected to sites $1$ and $N$.
Each bath is described by an independent (and infinite) set of bosonic operators $b_{\ell, n}$, with $n = 1,N$ and $\ell$ representing the internal indices of the bath modes. 
The Hamiltonian of each bath is
\begin{equation}\label{HB}
H_{B,n} =  \sum_\ell \Omega_\ell b_{\ell,n}^\dagger b_{\ell,n}
\end{equation}
where the bath frequencies $\Omega_\ell$ are assumed to take on a quasi-continuum of values. 
Moreover, the interaction Hamiltonian is taken to be linear in the system and bath operators:
\begin{equation}\label{Hint}
H_{I,n} = \sum\limits_\ell  c_{\ell,n} (a_n + a_n^\dagger) (b_{\ell,n} + b_{\ell,n}^\dagger),\qquad n = 1,N
\end{equation}
Thus, the composite system will evolve according to the total Hamiltonian 
\begin{equation}\label{Htot}
H_\text{tot} = H + H_{B,1} + H_{B,N} + H_{I,1} + H_{I,N}
\end{equation}

Throughout the process we assume that the baths remain in thermal equilibrium with a grand-canonical density matrices
\begin{equation}\label{bath_thermal}
\rho_{B,n} = \frac{e^{-\beta_n (H_{B,n} - \mu_n \mathcal{N}_{B,n})}}{Z_{B,n}}
\end{equation}
where $\beta_n$ and $\mu_n$ are the temperature and chemical potentials of each bath and $\mathcal{N}_{B,n} = \sum_\ell b_{\ell,n}^\dagger b_{\ell,n}$ is the bath number operator.  
Our goal is to trace out the bath degrees of freedom and obtain an effective Lindblad master equation for the chain, of the form 
\begin{equation}\label{master}
\frac{\ud \rho}{\ud t} = - i [H, \rho] + \mathcal{D}_1(\rho) + \mathcal{D}_N(\rho)
\end{equation}
where $\rho$ is the reduced density matrix of the system. 
Intuitively, one expects that $\mathcal{D}_1$ would act only on site $1$, and  $\mathcal{D}_N$ would act only on site $N$. 
However, we will show that when these dissipators are derived from a microscopic theory, they become non-local, affecting all normal modes of the system. 

The derivation of the effective Lindblad dissipators can be accomplished quite straightforwardly using the method of eigenoperators, discussed in detail in Ref.~\cite{Breuer2007}. 
Surprisingly, however, we are unaware of any publications in the literature which have applied this method to a many-body second-quantized system. 
We now briefly review its main features and then show how it can be applied to our present problem. 


\subsection{The method of eigenoperators}

Consider any physical system with Hamiltonian $H$, that is coupled to a bath via a Hamiltonian $H_I = A B$, where $A$ and $B$ are Hermitian system and bath operators respectively. 
In our case 
\begin{equation}\label{B_def}
B = \sum_\ell c_{\ell,n}(b_{\ell,n} + b_{\ell,n}^\dagger)
\end{equation} 
According to the method of eigenoperators, one may readily write down an effective Lindblad master equation for this bath interaction, assuming the usual Born-Markov and rotating wave approximations. 
The result is \cite{Breuer2007}:
\begin{equation}\label{D_gen}
D(\omega) = \sum\limits_\omega \Gamma(\omega) \bigg[ A(\omega) \rho A^\dagger(\omega) - \frac{1}{2} \{ A^\dagger(\omega) A(\omega), \rho\}\bigg]
\end{equation}
The different terms in this equation will now be explained in detail. 

Let $\epsilon$ and $\Pi_\epsilon$ denote the eigenvalues and eigen-projectors of the system Hamiltonian $H$. 
The operator $A(\omega)$ is called the eigenoperator of $A$ and is defined as 
\begin{equation}\label{eigen_op_def}
A(\omega) = \sum\limits_{\epsilon, \epsilon'} \Pi_\epsilon \; A \; \Pi_{\epsilon'} \; \delta_{\omega, \epsilon'-\epsilon}
\end{equation}
where $\omega$ therefore represents all allowed Bohr frequencies of the system. 
One may  verify that these operators satisfy the following relations: 
\begin{equation}\label{eigen_op_prop}
[H,A(\omega)] = - \omega A(\omega), \qquad A^\dagger(\omega) = A(-\omega)
\end{equation}
For many-body systems Eq.~(\ref{eigen_op_def}) is cumbersome to work with. 
Fortunately, all we will need is Eq.~(\ref{eigen_op_prop}), which we may also take as the definition of an eigenoperator. 

Next, we discuss the quantity $\Gamma(\omega)$ in Eq.~(\ref{D_gen}). 
It is defined as the Fourier transform of the bath correlation functions
\begin{equation}\label{Gamma1}
\Gamma(\omega) = \int\limits_{-\infty}^\infty e^{i \omega t} \langle B(t)  B(0) \rangle  \nonumber
\end{equation}
where $B(t) = e^{i H_B t} B e^{-i H_B t}$ is the Heisenberg representation of the bath operator $B$ and the expectation value in this formula is taken with respect to the bath thermal state~(\ref{bath_thermal}).
We may evaluate it explicitly for the case where $B$ is given in Eq.~(\ref{B_def}). 
For now we drop the indices $n$ since the derivation is the same for both baths. 
A straightforward calculation shows that $\Gamma$ may be written as 
\begin{equation}\label{Gamma2}
\Gamma(\omega) =  \begin{cases} 
\gamma(\omega) [1 + \bar{n}(\omega)], & \text{ if } \omega > 0	\\[0.2cm]
\gamma(-\omega) \bar{n}(-\omega), &	\text{ if } \omega < 0
\end{cases}
\end{equation}
where 
\begin{equation}\label{bose_einstein}
\bar{n}(\omega) = \frac{1}{e^{\beta(\omega- \mu)}-1}
\end{equation}
is the Bose-Einstein distribution and
\begin{equation}\label{spectral_density}
\gamma(\omega) = 2\pi \sum\limits_\ell c_\ell^2\; \delta(\omega - \Omega_\ell)
\end{equation}
is the spectral density of the system. 
Usually it is not possible to know the spectral density in detail, since it depends sensibly on the system bath couplings. 
The standard approach is to assume that it depends on $\omega$ as $\gamma(\omega) \sim \omega^\alpha$, up to a high cut-off and for some exponent $\alpha$. The case $\alpha = 1$ is usually referred to as an Ohmic spectral density.

\subsection{Derivation of the dissipators for the harmonic chain}

Now that we have a general recipe for constructing a dissipator, we may apply it to our specific problem. 
This amounts essentially to obtaining the eigenoperators $A_1(\omega)$ and $A_N(\omega)$ corresponding to $A_1 = a_1 + a_1^\dagger$ and $A_N = a_N + a_N^\dagger$. 
This is the main challenge of this method since, to find the eigenoperators one must know the entire eigen-structure of the Hamiltonian $H$ [Eq.~(\ref{H1})]. 
In our case, however, this task is straightforward since a quadratic Hamiltonian may always be diagonalized by a Bogoliubov transformation.
We first define a new set of bosonic operators $\eta_k$ according to 
\begin{equation}\label{a_eta}
a_n = \sum\limits_k S_{n,k} \;\eta_k
\end{equation}
where
\begin{equation}\label{S}
S_{n,k} = \sqrt{\frac{2}{N+1}} \sin(nk), \quad k = \frac{\pi}{N+1}, \ldots, \frac{N\pi}{N+1}
\end{equation}
is the Fourier sine transform matrix. 
The unitary nature of the $S_{n,k}$ preserves the bosonic algebra of the $\eta_k$. 
In terms of these new operators  the Hamiltonian~(\ref{H1}) becomes  
\begin{equation}\label{H2}
H = \sum\limits_k E_k \;\eta_k^\dagger \eta_k,\qquad E_k = \epsilon - g \cos k
\end{equation}
This result shows clearly that the model will only have a stable ground-state for $\epsilon>|g|$. 

To find the eigenoperators $A_{1,N}(\omega)$ we now note that the $\eta_k$ 
satisfy $[H,\eta_k] = - E_k \eta_k$. 
Comparing this with Eq.~(\ref{eigen_op_prop}) then shows that $\eta_k$ is itself an eigenoperator of $H$ with Bohr frequency $E_k$. 
Similarly, $\eta_k^\dagger$ will be an eigenoperator with frequency $-E_k$. 
The eigenoperator $A_1(\omega)$ corresponding to the operator $A_1 = a_1+a_1^\dagger$ will then be 
\begin{equation}\label{eigen_A1}
A_1(\omega) = \sum\limits_k S_{1,k} \bigg[ \eta_k \;\delta_{\omega, E_k} + \eta_k^\dagger \; \delta_{\omega, -E_k}\bigg]
\end{equation}
This can also be shown using Eq.~(\ref{eigen_op_def})  directly. 
However, this approach is much more cumbersome and we prefer to use  Eq.~(\ref{eigen_op_prop}).
Similarly, the eigenoperator $A_N(\omega)$ corresponding to the operator $A_N = a_N+a_N^\dagger$ will be
\begin{equation}\label{eigen_AN}
A_N(\omega) = \sum\limits_k S_{N,k} \bigg[ \eta_k \;\delta_{\omega, E_k} + \eta_k^\dagger \; \delta_{\omega, -E_k}\bigg]
\end{equation}
As a sanity check, if $g = 0$ then $E_k = \epsilon$ and the $\delta$'s may be taken out of Eqs.~(\ref{eigen_A1}) and (\ref{eigen_AN}). 
Using Eq.~(\ref{a_eta}) we then get 
\begin{equation}\label{eigen_A1_2}
A_1(\omega) = a_1\; \delta_{\omega,\epsilon} + a_1^\dagger\; \delta_{\omega,-\epsilon}\qquad (g \to 0)
\end{equation} 
and similarly for $A_N$. 
Physically, this means that when $g=0$ the only allowed transitions that can be produced by $(a_1+a_1^\dagger)$ are those which affect only the first mode. Conversely, when $g\neq 0$ then $(a_1+a_1^\dagger)$ may cause transitions which influence all normal modes of the system.

The next step is to substitute each of these terms into Eq.~(\ref{D_gen}) to find the corresponding dissipator. 
In order to simplify the formulas, let us look at a typical term like $A_1(\omega) \rho A_1^\dagger (\omega)$. 
In doing so, we must keep in mind that the single-particle energy eigenvalues $E_k$ are non-negative and non-degenerate. We then get:
\begin{widetext}
\begin{IEEEeqnarray*}{rCl}
\sum\limits_\omega \Gamma_1(\omega)  A_1(\omega) \rho A_1^\dagger (\omega)
&=& \sum\limits_{\omega,k,q} \Gamma_1(\omega) S_{1,k} S_{1,q}
(\eta_k \;\delta_{\omega, E_k} + \eta_k^\dagger \; \delta_{\omega, -E_k}) \rho 
(\eta_q^\dagger \;\delta_{\omega, E_q} + \eta_q \; \delta_{\omega, -E_q})
\\[0.2cm]
&=& \sum\limits_{\omega,k,q} \Gamma_1(\omega) S_{1,k}S_{1,q} (\eta_k \rho \eta_q^\dagger \; \delta_{\omega,E_k} \delta_{\omega,E_q} + \eta_k^\dagger \rho \eta_q \;\delta_{\omega,-E_k}\delta_{\omega,-E_q})
\\[0.2cm]
&=& \sum\limits_k   S_{1,k}^2\bigg[ \Gamma_1(E_k) \; \eta_k \rho \eta_k^\dagger  + \Gamma_1(-E_k) \eta_k^\dagger \rho \eta_k\bigg]
\end{IEEEeqnarray*}
A similar structure will follow for all other terms in the dissipator. 
Finally, substituting Eq.~(\ref{Gamma2}) for $\Gamma(E_k)$ and $\Gamma(-E_k)$, we arrive at the dissipators
\begin{equation}\label{Dn}
\mathcal{D}_n(\rho) = \sum\limits_k S_{n,k}^2 \gamma_n(E_k) \Bigg\{ 
[1 + \bar{n}_{n,k}] \bigg[ \eta_k\rho\eta_k^\dagger - \frac{1}{2}\{\eta_k^\dagger \eta_k, \rho\}\bigg]
+   \bar{n}_{n,k}
\bigg[ \eta_k^\dagger\rho\eta_k - \frac{1}{2}\{\eta_k \eta_k^\dagger, \rho\}\bigg]\Bigg\}
\end{equation}

where $n = 1,N$ and 
\begin{equation}\label{bose_einstein2}
\bar{n}_{n,k} = \frac{1}{e^{\beta_n(E_k -\mu_n)}-1}
\end{equation}
is the Bose-Einstein distribution describing the contact of mode $k$ with  bath $n = 1,N$. 
This concludes our derivation of the effective dissipator corresponding to the bath interactions~(\ref{Hint}).
To our knowledge, this is the first microscopic derivation of an effective Lindblad dissipator for a many-body system.

Eq.~(\ref{Dn}) has several points worth discussing. 
We began with a microscopic theory where only two sites were coupled to the bath. 
However, we see here that since the different sites interact, this coupling affects all normal modes of the system. 
This is a global behavior and is expected for any real heat bath.  
In the limit where the site interaction $g$ tends to zero, we may find the dissipator by using instead Eq.~(\ref{eigen_A1_2}). Retracing the same steps as above we then find the local dissipators
\begin{equation}
\mathcal{D}_{n,\text{local}}(\rho) = \gamma_n (1+\bar{n}_n) \bigg[a_n \rho a_n^\dagger - \frac{1}{2}\{a_n^\dagger a_n, \rho\}\bigg] + \gamma_n \bar{n}_n \bigg[a_n^\dagger \rho a_n - \frac{1}{2}\{a_n a_n^\dagger,\rho\}\bigg]
\end{equation}
\end{widetext}
This is precisely the dissipators studied in Ref.~\cite{Asadian2013}. 
We therefore expect that all results that will be derive below should tend to those of Ref.~\cite{Asadian2013} in the limit $g/\epsilon \ll 1$
\footnote{In order to make this comparison, however, there is a subtlety concerning the relative magnitudes of $g$ and the spectral density $\gamma$. The entire theory of Lindblad dissipators is based on the assumption of weak coupling, which means $\gamma$ should be much smaller than any energy scale of the system. The correct limit for comparing the two models is, therefore, $\gamma \ll g \ll \epsilon$. In this limit all results of the present paper tend to those of Ref.~\cite{Asadian2013}.  }.

\begin{figure}[!h]
\centering
\includegraphics[width=0.22\textwidth]{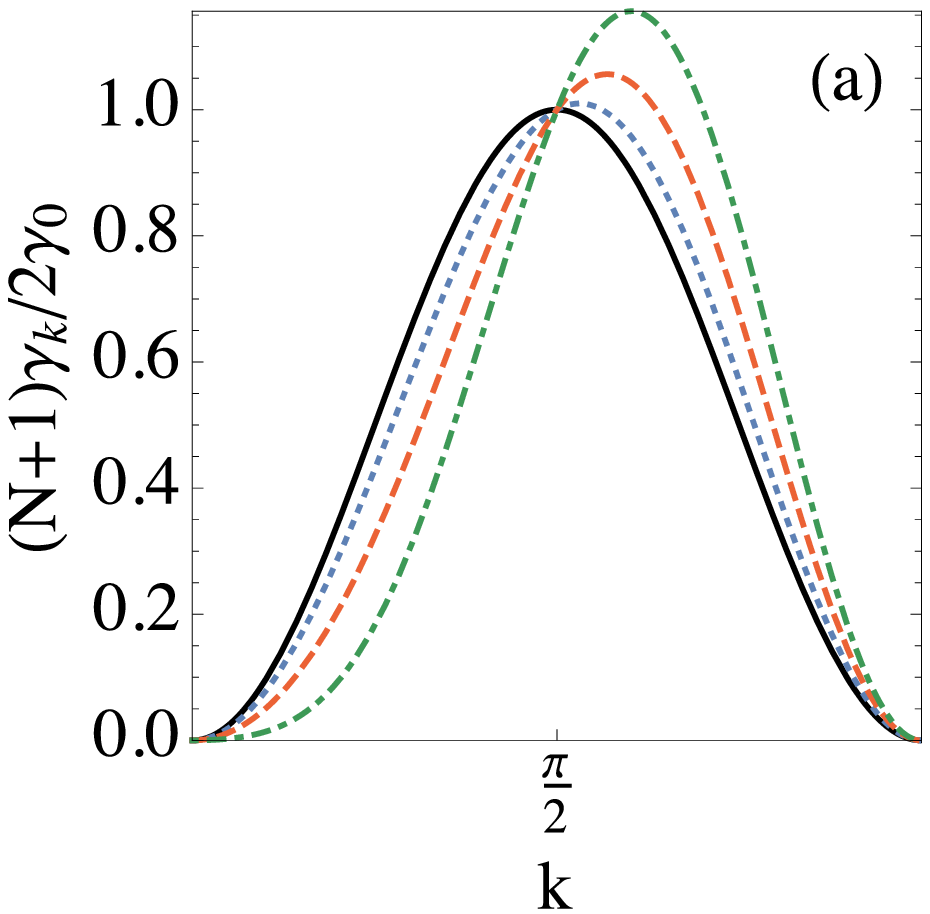}\quad
\includegraphics[width=0.22\textwidth]{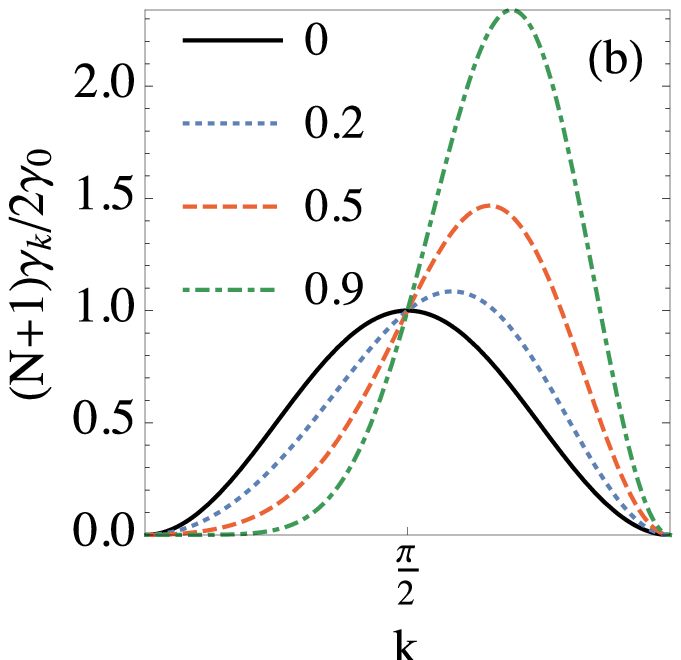}
\caption{\label{fig:spectral_density}
The effective system bath coupling $\gamma_k$ in Eq.~(\ref{gamma_effective}) for 
(a) $\alpha = 1$ and (b) $\alpha = 3$. 
Each image contains curves for different values of $g/\epsilon$ [as labeled in image (b)].
}
\end{figure}

Moreover, we also see in Eq.~(\ref{Dn}) that the information about which particle was initially coupled to the bath is reduced only to   the functions $S_{n,k}^2$, which enter as an effective system-bath coupling $\gamma_{n,k} = S_{n,k}^2 \gamma(E_k)$ for each mode $k$. 
In the derivation nowhere have we used the specific value of these functions. Thus, one may also contemplate other types of tight-binding models such as, for instance, a disordered chain, where the tunneling constants $g_n$ become inhomogeneous.
In our case, due to the symmetry of the problem, it follows from Eq.~(\ref{S}) that $S_{1,k}$ and $S_{N,k}$ differ by at most a minus sign. 
Consequently, 
\begin{equation}
S_{1,k}^2 = S_{N,k}^2 = \frac{2}{N+1} \sin^2 k
\end{equation}
The effective coupling of mode $k$ to the heat bath is therefore 
$\gamma_{n,k} =\frac{2}{N+1} \sin^2 k \gamma_n(E_k)$.
As discussed above, we will assume for concreteness that  the spectral density  may be modeled as $\gamma_n(\omega) = \gamma_{n,0} \omega^{\alpha_n}$, for some exponent $\alpha_n$ for each bath. 
Thus we will henceforth assume that 
\begin{equation}\label{gamma_effective}
\gamma_{n,k} =\gamma_{n,0} \frac{2}{N+1} \sin^2 k( \epsilon - g/\epsilon \cos k)^{\alpha_n}
\end{equation}
where the factor of $\epsilon$ in the denominator was adjusted simply so that $\gamma_{n,0}$ continues to have units of frequency. 
This function is illustrated in Fig.~\ref{fig:spectral_density} for different values of $g$ and $\alpha$. 
As can be seen, the coupling of the modes with $k\sim 0$ and $k \sim \pi$ tends to zero. 
Moreover, a higher  value of $\alpha$ introduces an asymmetry in the couplings. 
The mode occupations $\langle \eta_k^\dagger \eta_k \rangle$ usually relax toward thermal equilibrium proportionally to $e^{-\gamma_k t}$. 
Thus, the results in Fig.~\ref{fig:spectral_density} illustrate  the different relaxation time scales of the normal modes. 

Irrespective of the value of $\gamma_k$, when $T_1 = T_2 = T$ and $\mu_1 = \mu_2 = \mu$,  the master Eq.~(\ref{master}) with the microscopic dissipators~(\ref{Dn}) will move the system toward the true thermal Gibbs state
\begin{equation}
\rho_\text{eq} = \prod\limits_k \frac{ e^{-\beta (E_k-\mu) \; \eta_k^\dagger \eta_k}}{Z}
\end{equation}
for any initial condition.
Moreover, it can be shown \cite{Breuer2007}  that dissipators derived using the method of eigenoperators will satisfy detailed balance. 
This is actually a consequence of the Kubo-Martin-Schinger relation, which implies that $\Gamma(\omega)$ in Eq.~(\ref{Gamma2}) will satisfy 
\begin{equation}
\Gamma(-\omega) = e^{-\beta(\omega-\mu)} \Gamma(\omega)
\end{equation}
A correct thermal target state and a relaxation dynamics which obeys detailed balance   are the two most fundamental properties one expects from a physical model of the interaction with a  heat bath. 
Hence, one expects  that these dissipators should provide an accurate modeling of the system bath interaction. 
The results which will be discussed below corroborate this expectation.

%
%
%
%
\section{\label{sec:ss}Properties of the steady-state}

\subsection{\label{ssec:currents}Particle and energy currents}

We now study the non-equilibrium steady-state (NESS) produced by the master Eq.~(\ref{master}).
As the first step, we establish formulas for the particle and energy currents in the system. 

Let 
\begin{equation}
\mathcal{N} = \sum\limits_n a_n^\dagger a_n = \sum\limits_k \eta_k^\dagger \eta_k
\end{equation}
be the total number of particles in the system. 
Using the master Eq.~(\ref{master}) and noting that $[H,\mathcal{N}] = 0$, we find 
\begin{IEEEeqnarray}{rCl}
\frac{\ud \langle \mathcal{N} \rangle}{\ud t} &=&  \tr\bigg\{ \mathcal{N} \mathcal{D}_1(\rho) \bigg\} 
+  \tr\bigg\{ \mathcal{N} \mathcal{D}_N(\rho) \bigg\} 
\nonumber
\\[0.2cm]
&:=& J_{\mathcal{N}_1} - J_{\mathcal{N}_N}
\end{IEEEeqnarray}
which define the currents $J_{\mathcal{N}_1}$ and $J_{\mathcal{N}_N}$,  of particles entering site 1 toward the left bath and leaving site $N$ toward the right bath.
In the steady-state $\ud \langle \mathcal{N} \rangle/\ud t = 0$ so that all particles entering site $1$ eventually leave site $N$:
\begin{equation}
J_{\mathcal{N}_1}  = J_{\mathcal{N}_N} := \JN
\end{equation}
Using Eq.~(\ref{Dn}) for $\mathcal{D}_1$ (or $\mathcal{D}_N$) we find that $\JN$ may be written as 
\begin{equation}\label{JN_def}
\JN = \sum\limits_k \gamma_{1,k} (\bar{n}_{1,k} - \langle \eta_k^\dagger \eta_k \rangle) = 
-\sum\limits_k \gamma_{N,k} (\bar{n}_{N,k} - \langle \eta_k^\dagger \eta_k \rangle) 
\end{equation}
which is a convenient formula for the particle current. 

We may proceed similarly with the energy current:
\begin{IEEEeqnarray}{rCl}
\frac{\ud \langle H \rangle}{\ud t} &=&  \tr\bigg\{ H \mathcal{D}_1(\rho) \bigg\} 
+  \tr\bigg\{ H \mathcal{D}_N(\rho) \bigg\} 
\nonumber
\\[0.2cm]
&:=& J_{\mathcal{E}_1} - J_{\mathcal{E}_N}
\end{IEEEeqnarray}
In the steady-state  
\begin{equation}
J_{E_1}  = J_{E_N} := \JE
\end{equation}
Using Eqs.~(\ref{H2}) and (\ref{Dn}) we then find 
\begin{IEEEeqnarray}{rCl}\label{JE_def}
J_E &=& \sum\limits_k \gamma_{1,k} E_k (\bar{n}_{1,k} - \langle \eta_k^\dagger \eta_k \rangle) \\[0.2cm]
\nonumber &=& 
-\sum\limits_k \gamma_{N,k} E_k (\bar{n}_{N,k} - \langle \eta_k^\dagger \eta_k \rangle) 
\end{IEEEeqnarray}

\subsection{\label{ssec:occu}Steady-state occupation numbers}

The time evolution of the correlation functions $\langle \eta_k^\dagger \eta_q \rangle$ may be readily obtained  from the master Eq.~(\ref{master}). They read
\begin{IEEEeqnarray}{rCl}
\frac{\ud \langle \eta_k^\dagger \eta_q \rangle}{\ud t} &=&-
\frac{(\gamma_{1,k}+\gamma_{N,k} + \gamma_{1,q}+\gamma_{N,q})}{2} \langle \eta_k^\dagger \eta_q \rangle
\\[0.2cm]
\nonumber
&&\qquad+ \delta_{k,q}( \gamma_{1,k} \bar{n}_{1,k} + \gamma_{N,k} \bar{n}_{N,k})
\end{IEEEeqnarray}
We therefore see that, in the long-time limit the off-diagonal terms $\langle \eta_k^\dagger \eta_q \rangle$ (with $q\neq k$) will vanish, whereas the diagonal terms will tend to 
\begin{equation}\label{occupations}
\langle \eta_k^\dagger \eta_k \rangle = \frac{\gamma_{1,k} \bar{n}_{1,k} + \gamma_{N,k} \bar{n}_{N,k}}{\gamma_{1,k} + \gamma_{N,k}} 
\end{equation}

The occupation numbers in coordinate space are obtained from Eq.~(\ref{a_eta}) and read 
\begin{IEEEeqnarray}{rCl}
\nonumber
\langle a_n^\dagger a_n \rangle &=& \sum\limits_{k,k'} S_{n,k} S_{n,k'} \langle \eta_k^\dagger \eta_{k'}\rangle \\[0.2cm]
&=& \frac{2}{N+1}\sum\limits_k \sin^2(n k) \frac{\gamma_{1,k} \bar{n}_{1,k} + \gamma_{N,k} \bar{n}_{N,k}}{\gamma_{1,k} + \gamma_{N,k}} 
\end{IEEEeqnarray}
If we assume that $g/\epsilon \ll 1$ then the energy levels become  roughly independent of $k$:  $E_k = \epsilon - g \cos k\sim \epsilon$.
Consequently, so will  $\bar{n}_{n,k}$ and $\gamma_{n,k}$, which  may thus be taken outside  the sum. 
The resulting sum is $\sum_k \sin^2(nk) = (N+1)/2$. Thus, we conclude that in the limit $g/\epsilon \ll 1$, the coordinate space occupations 
will tend to 
\begin{equation}
\langle a_n^\dagger a_n \rangle \simeq \frac{\gamma_1 \bar{n}_1 + \gamma_N \bar{n}_N}{\gamma_1 +\gamma_N}
\end{equation}
which is a simple arithmetic average of the bath occupations. 
This result agrees with the calculations in Ref.~\cite{Asadian2013}.

Substituting Eq.~(\ref{occupations}) in Eqs.~(\ref{JN_def}) and (\ref{JE_def}) we get for the particle and energy currents
\begin{IEEEeqnarray}{rCl}
\JN &=& \sum\limits_k \frac{\gamma_{1,k} \gamma_{N,k}}{\gamma_{1,k} + \gamma_{N,k}} (\bar{n}_{1,k} - \bar{n}_{N,k})	\\[0.2cm]
\JE &=& \sum\limits_k \frac{\gamma_{1,k} \gamma_{N,k}}{\gamma_{1,k} + \gamma_{N,k}} E_k (\bar{n}_{1,k} - \bar{n}_{N,k})
\end{IEEEeqnarray}
When $\gamma_{1,k} = \gamma_{N,k} = \gamma_k$ this simplifies to 
\begin{IEEEeqnarray}{rCl}
\JN &=& \frac{1}{2}\sum\limits_k \gamma_k (\bar{n}_{1,k} - \bar{n}_{N,k})	\\[0.2cm]
\JE &=& \frac{1}{2}\sum\limits_k \gamma_k  E_k (\bar{n}_{1,k} - \bar{n}_{N,k})
\end{IEEEeqnarray}
The currents therefore are seen to have the structure of the Landauer-B\"utikker formula \cite{Landauer1987,VanWees1988,Pastawski1991,Baringhaus2014,Datta1997}, with the spectral density $\gamma_k$ playing the role of the transfer integral; i.e., of the probability to observe a tunneling of an excitation from the bath towards the system, an interpretation which agrees intuitively with the basic structure of the system-bath interaction~(\ref{Hint}).
Putting it differently, the heat baths play the role of the leads and the chain, being harmonic, functions as a perfectly conducting channel through which the excitations may flow.

In the limit $g/\epsilon \ll 1$ the occupation numbers become independent of $k$ and we get, using the same arguments as above,  
\begin{IEEEeqnarray}{rCl}
\JN &=& \frac{\gamma_0}{2} (\bar{n}_{1} - \bar{n}_{N})	\\[0.2cm]
\JE &=& \frac{\gamma_0}{2} \epsilon\; (\bar{n}_{1} - \bar{n}_{N})
\end{IEEEeqnarray}
These results again coincide  with those of Ref.~\cite{Asadian2013}.  

\subsection{\label{ssec:thermo}Thermodynamic limit}

In the thermodynamic limit we may replace the sum with an integral using the recipe $\sum_k \to \frac{N}{\pi} \int_0^\pi \ud k$. 
Using also Eq.~(\ref{gamma_effective}) to substitute for $\gamma_k$, we then find 
\begin{IEEEeqnarray}{rCl}
\JN &=& \frac{\gamma_0}{\pi} \int\limits_0^\pi \ud k \; \sin^2 k(1- g/\epsilon \cos k)^\alpha (\bar{n}_{1,k} - \bar{n}_{N,k})
\IEEEeqnarraynumspace	\\[0.2cm]
\JE &=& \frac{ \gamma_0}{\pi} \int\limits_0^\pi \ud k \; \sin^2 k(1- g/\epsilon \cos k)^{\alpha}E_k (\bar{n}_{1,k} - \bar{n}_{N,k})
\IEEEeqnarraynumspace
\end{IEEEeqnarray}
We may also assume infinitesimal temperature and chemical potential imbalances; that is, we choose 
$T_1 = T+ \Delta T/2$, $T_2 = T - \Delta T/2$, $\mu_1 = \mu + \Delta\mu/2$ and $\mu_2 = \mu - \Delta\mu/2$, where $\Delta T$ and $\Delta \mu$ are assumed to be small quantities. 
This allow us to write
\begin{equation}
\bar{n}_{1,k} - \bar{n}_{N,k} = \frac{\partial \bar{n}_k}{\partial T}\Delta T + \frac{\partial \bar{n}_k}{\partial \mu} \Delta \mu
\end{equation}
where $\bar{n}_k = (e^{\beta(E_k -\mu)}-1)^{-1}$. 
We then get 
\begin{IEEEeqnarray}{rCl}
\label{JN2}
\JN &=& \frac{\partial \FN}{\partial T}\Delta T + \frac{\partial \FN}{\partial \mu} \Delta \mu 	\\[0.2cm]
\label{JE2}
\JE &=& \frac{\partial \FE}{\partial T}\Delta T + \frac{\partial \FE}{\partial \mu} \Delta \mu
\end{IEEEeqnarray}
where 
\begin{IEEEeqnarray}{rCl}
\label{FN}
\FN &=&  \frac{\gamma_0}{\pi} \int\limits_0^\pi \ud k \; \sin^2 k(1- g/\epsilon \cos k)^\alpha \bar{n}_k	\\[0.2cm]
\label{FE}
\FE &=& \frac{\gamma_0}{\pi} \int\limits_0^\pi \ud k \; \sin^2 k(1- g/\epsilon \cos k)^{\alpha} E_k \bar{n}_k
\end{IEEEeqnarray}

\begin{figure}
\centering
\includegraphics[width=0.45\textwidth]{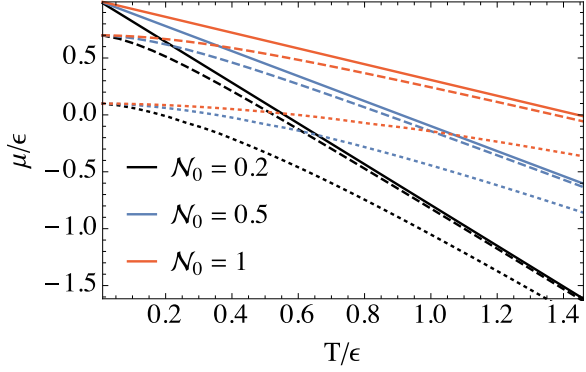}
\caption{\label{fig:chem}The chemical potential $\mu$ as a function of temperature, computed numerically from Eq.~(\ref{N_ss}), for different values of  $\mathcal{N} = \mathcal{N}_0$ and with $g/\epsilon = 0$ (solid curves), $g/\epsilon = 0.3$ (dashed curves) and $g/\epsilon = 0.9$ (dotted curvers). 
}
\end{figure}

It is also worth noting that, for these infinitesimal imbalances in $T$ and $\mu$, the total number of particles becomes, up to terms quadratic in $\Delta T$ and $\Delta \mu$, 
\begin{equation}\label{N_ss}
\langle \mathcal{N} \rangle = \sum\limits_{k} \bar{n}_k = \frac{N}{\pi} \int\limits_0^\pi \frac{\ud k}{e^{\beta(E_k-\mu)}-1}
\end{equation}
This equation may be used to fix the average chemical potential $\mu$ of the two baths in such a way that the total number of particles in the chain remains fixed at a given value $\langle \mathcal{N} \rangle = \mathcal{N}_0$. 
In other words, for small imbalances (linear response) one may study chemical potential gradients while keeping the number of particles in the chain fixed. 
The chemical potential as a function of $T$ is illustrated in Fig.~\ref{fig:chem} for different choices of $g$ and $\mathcal{N}_0$ (with all energies measured in unites of $\epsilon$).
For 1D there is no Bose-Einstein condensation, except at $T\to0$,  where the chemical potential tends to $\epsilon -g$. 
Moreover, when $g \to 0$ the integral in Eq.~(\ref{N_ss}) becomes independent of $k$ and we obtain 
\begin{equation}
\epsilon-\mu = T \ln \left(\frac{N+\mathcal{N}_0}{\mathcal{N}_0}\right)
\end{equation}
which shows that $\mu$ decreases linearly with increasing temperature. 
When $g\neq 0$ this linear behavior is bent, as seen in Fig.~\ref{fig:chem}.

\begin{figure}
\centering
\includegraphics[width=0.22\textwidth]{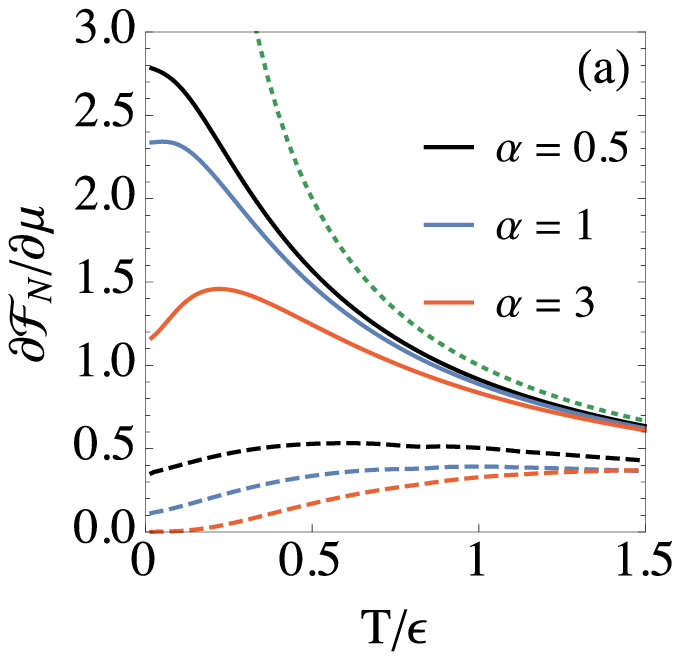}\quad
\includegraphics[width=0.22\textwidth]{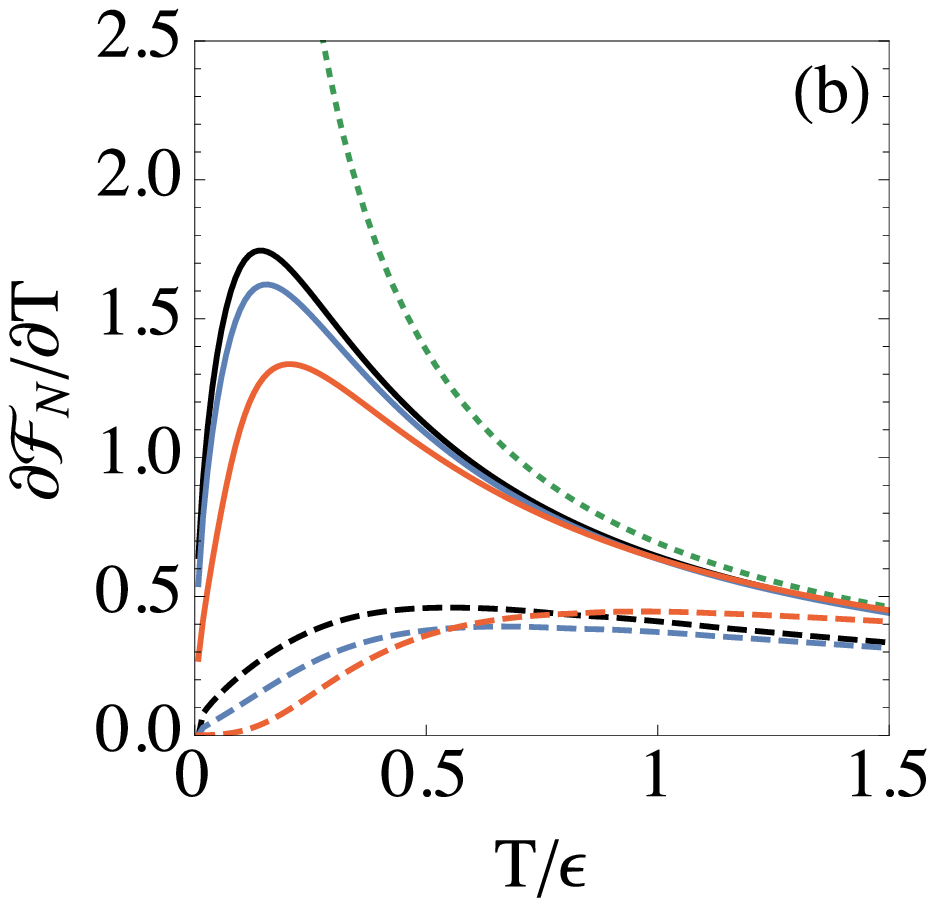}\\
\includegraphics[width=0.22\textwidth]{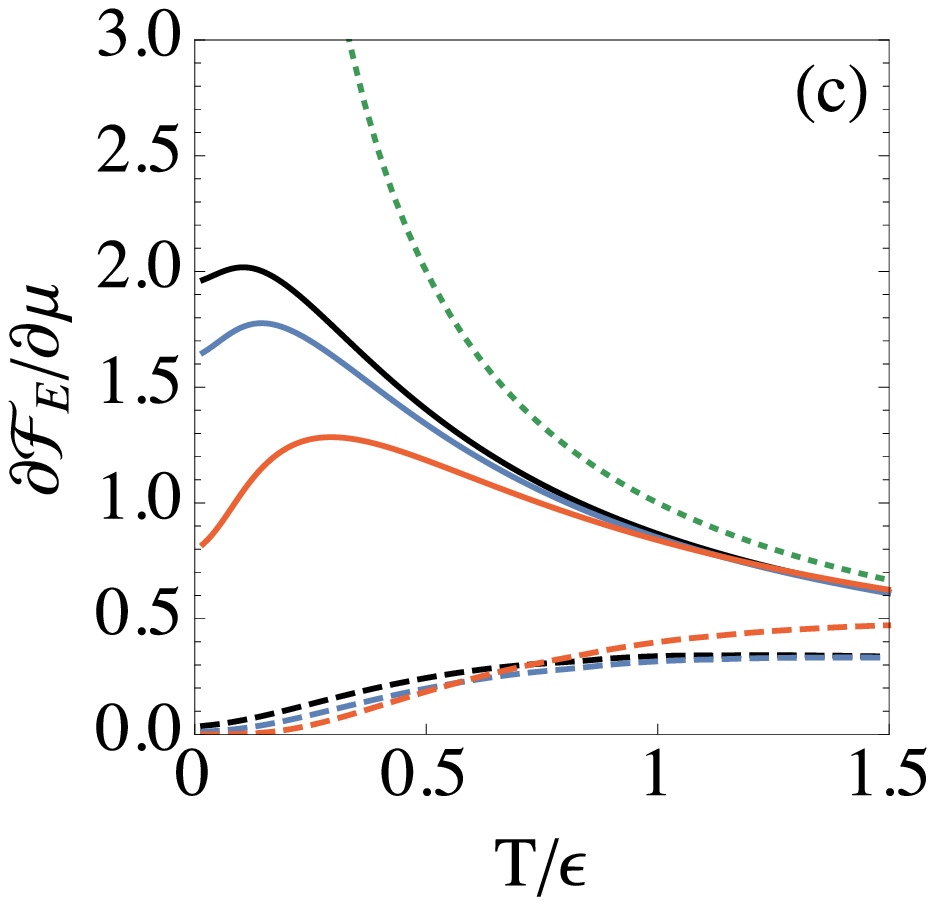}\quad
\includegraphics[width=0.22\textwidth]{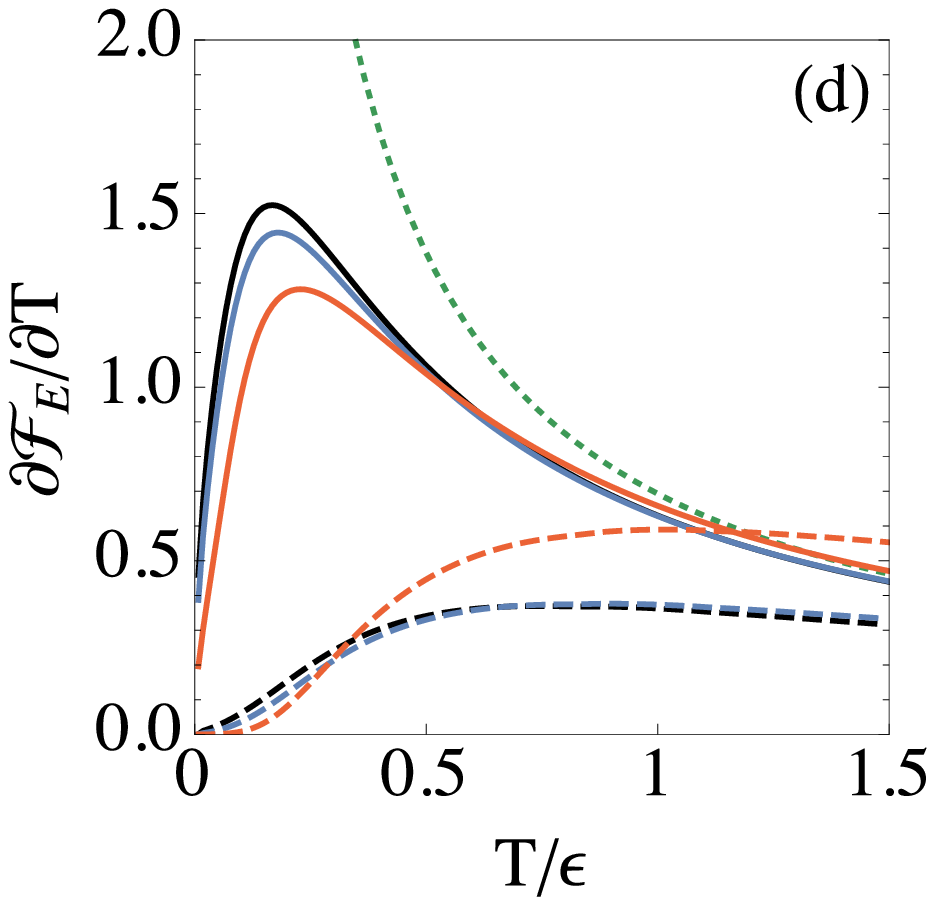}
\caption{\label{fig:fluxes}The different contributions to the particle and energy fluxes, Eqs.~(\ref{JN2}) and (\ref{JE2}) as a function of temperature for different values of $\alpha$ [as shown in image (a)] and $g/\epsilon = 0.3$ (solid curves) and $g/\epsilon = 0.9$ (dashed curves). The mean occupation number is fixed at $\mathcal{N}_0/N = 1$ and all curves are given in units of $\gamma_0=\epsilon=1$. 
The dotted green line denote the currents in the limit $g/\epsilon \ll 1$. 
}
\end{figure}

In Fig.~\ref{fig:fluxes} we present the behavior of the four contributions, $\partial \FN/\partial \mu$, $\partial \FN/\partial T$, $\partial \FE/\partial \mu$ and $\partial \FE/\partial T$ to the currents in  Eqs.~(\ref{JN2}) and (\ref{JE2}), assuming a fixed average number of particles $\mathcal{N}_0$ in the chain. The curves are for  different values of $\alpha$ [cf. Eq.~(\ref{gamma_effective})] and $g$, with fixed $\mathcal{N}_0/N = 1$. 
The currents in the limit  $g/\epsilon \ll 1$ are depicted by dotted green curves in each figure.

\subsection{\label{ssec:onsager}Onsager coefficients}

\begin{figure}[!h]
\centering
\includegraphics[width=0.22\textwidth]{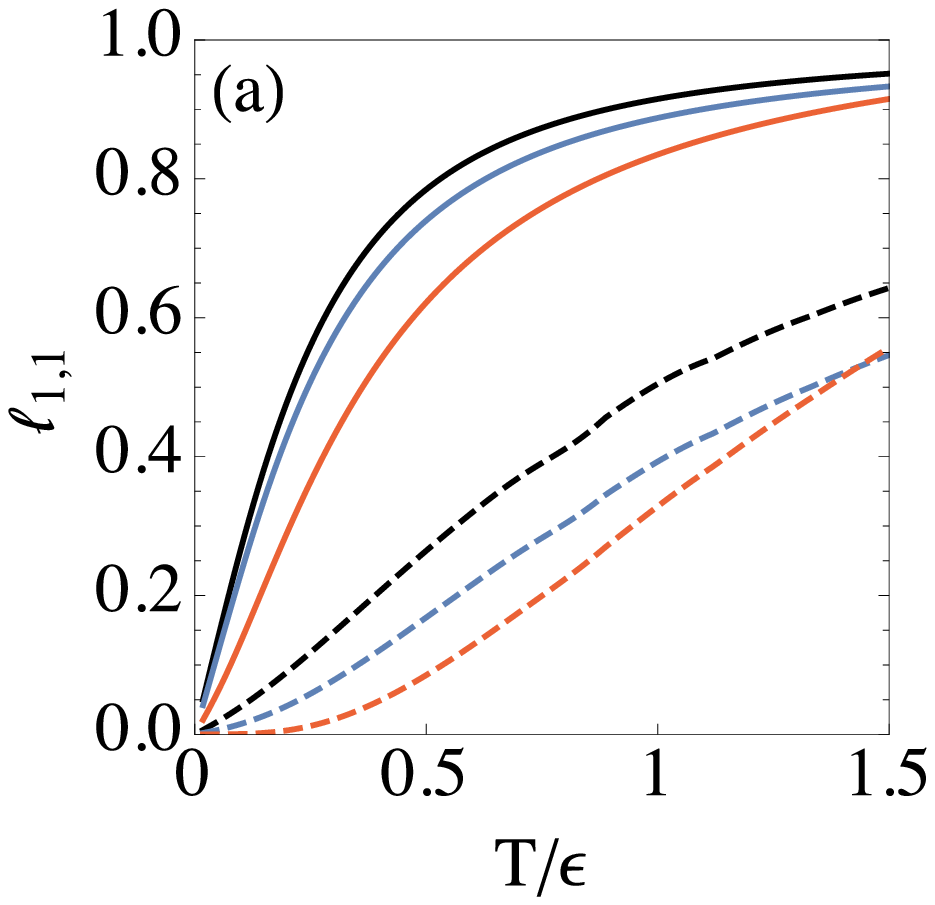}\quad
\includegraphics[width=0.22\textwidth]{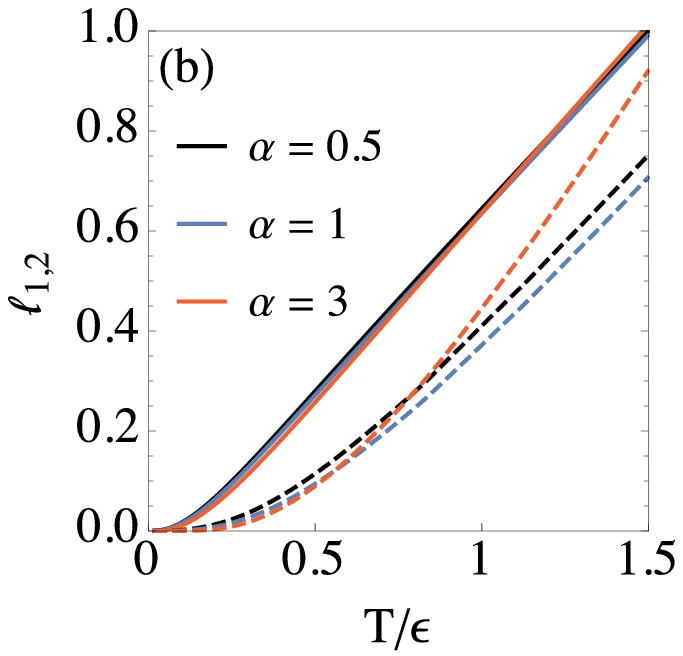}\\
\includegraphics[width=0.22\textwidth]{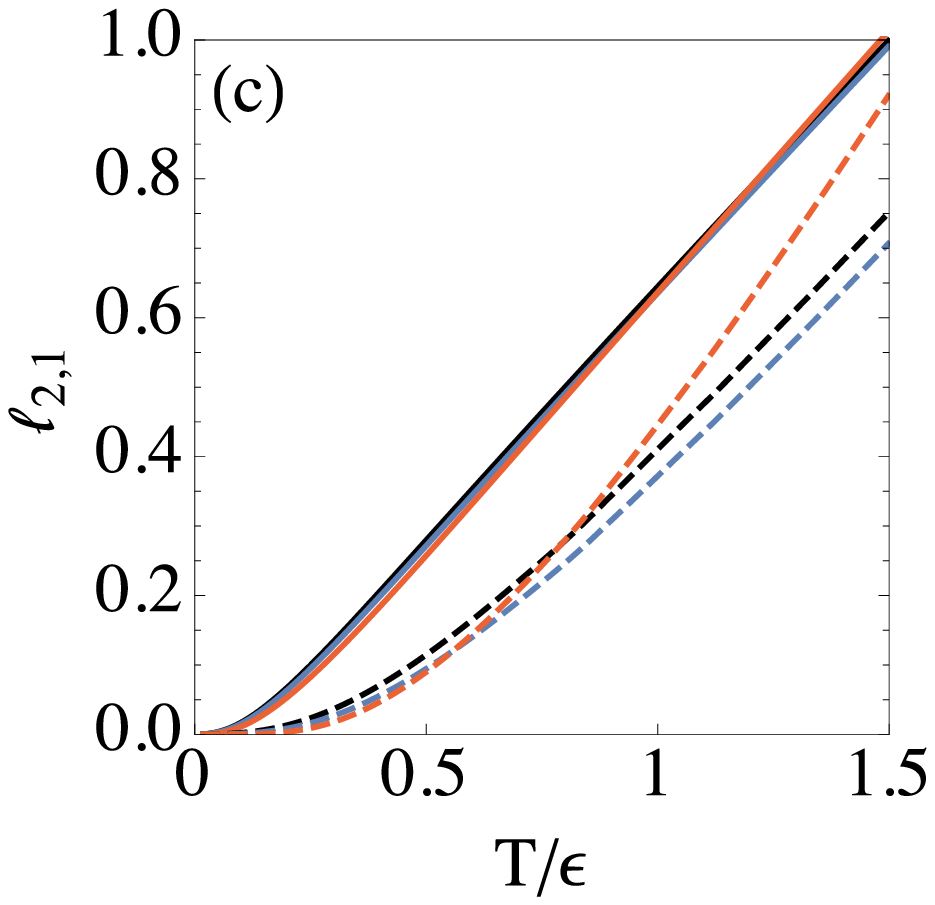}\quad
\includegraphics[width=0.22\textwidth]{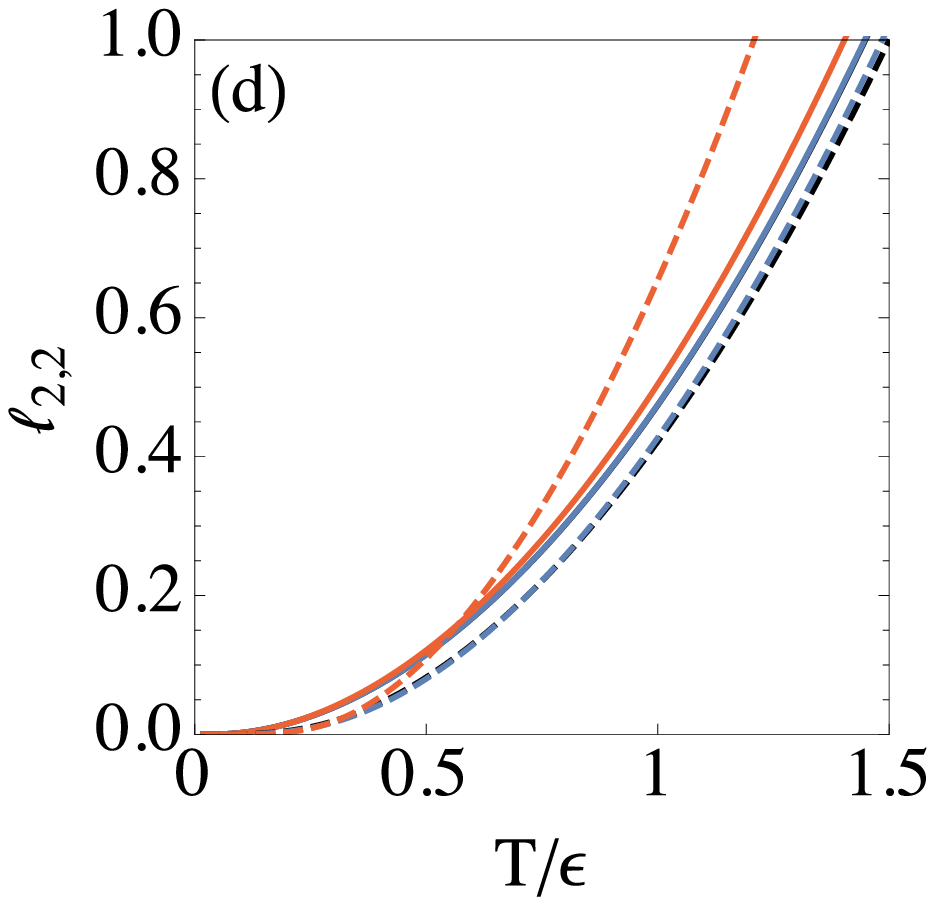}
\caption{\label{fig:onsager} 
The Onsager coefficients, Eqs.~(\ref{l11})-(\ref{l22}) for the same conditions as Fig.~\ref{fig:fluxes}.
}
\end{figure}

According to the first law, the energy current $\JE$ may be decomposed into a heat current $\JQ$ and a particle current $\mu\JN$.
This can be used to define the heat current through the system as $\JQ = \JE - \mu \JN$ \cite{Mahan}. 
Following Onsager \cite{Onsager1931a,Onsager1931} we now relate the particle and heat currents to the generalized forces $\Delta\mu/T$ and $\Delta T/T^2$:
\begin{IEEEeqnarray}{rCl}
\JN &=& \ell_{1,1} \frac{\Delta \mu}{T} + \ell_{1,2} \frac{\Delta T}{T^2}\\[0.2cm]
\JQ &=& \ell_{2,1} \frac{\Delta \mu}{T} + \ell_{2,2} \frac{\Delta T}{T^2}
\end{IEEEeqnarray}
The Onsager coefficients $\ell_{i,j}$ may be read off directly from Eqs.~(\ref{JN2}) and (\ref{JE2}):
\begin{IEEEeqnarray}{rCl}
\label{l11}
\ell_{1,1} &=& T \frac{\partial \FN}{\partial \mu}\\[0.2cm]
\label{l12}
\ell_{1,2} &=& T^2 \frac{\partial \FN}{\partial T}\\[0.2cm]
\label{l21}
\ell_{2,1} &=& T \bigg[ \frac{\partial \FE}{\partial \mu} - \mu\frac{\partial \FN}{\partial \mu}\bigg]\\[0.2cm]
\label{l22}
\ell_{2,2} &=& T^2 \bigg[ \frac{\partial \FE}{\partial T} - \mu \frac{\partial \FN}{\partial T}\bigg]
\end{IEEEeqnarray}
These coefficients are illustrated in Fig.~\ref{fig:onsager} for the same conditions as Fig.~\ref{fig:fluxes}.

It is also possible to verify that, since 
\[
T \frac{\partial \bar{n}_k}{\partial T} = (E_k -\mu) \frac{\partial \bar{n}_k}{\partial \mu}
\]
it follows that 
\begin{equation}
\ell_{1,2} = \ell_{2,1}
\end{equation}
which is Onsager's reciprocity relation. 
This is an important result. 
It corroborates the consistency of the non-equilibrium behavior generated by the microscopic Lindblad dissipators~(\ref{Dn}).
Of course, this result is also expected in view of the fact that these dissipators satisfy detailed balance, which is the primary physical basis for Onsager's relation.

\begin{figure}
\centering
\includegraphics[width=0.45\textwidth]{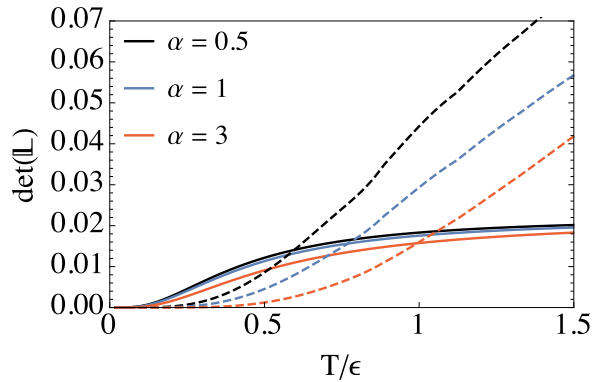}
\caption{\label{fig:onsager_det}
The determinant of the Onsager matrix $\det(\mathbb{L}) = \ell_{1,1} \ell_{2,2} - \ell_{1,2} \ell_{2,1}$ for the same conditions as Figs.~\ref{fig:fluxes} and \ref{fig:onsager}. 
}
\end{figure}

The entropy production rate in the system is then given by the quadratic form
\begin{equation}
\Pi = \begin{pmatrix}\frac{\Delta\mu}{T} & \frac{\Delta T}{T^2} \end{pmatrix}
\begin{pmatrix}
\ell_{1,1} & \ell_{1,2} \\[0.2cm]
\ell_{2,1} & \ell_{2,2} 
\end{pmatrix}
\begin{pmatrix}\frac{\Delta\mu}{T} \\[0.2cm] \frac{\Delta T}{T^2} \end{pmatrix}
\end{equation}
Its non-negativity will be ensured provided  the Onsager matrix $\mathbb{L}$ is positive semi-definite. 
This can be examined by looking at the non-negativity of its determinant $\det(\mathbb{L}) = \ell_{1,1} \ell_{2,2} - \ell_{1,2} \ell_{2,1}$.
This is illustrated in Fig.~\ref{fig:onsager_det}, where the non-negativity is manifested. 
We also see in this figure a non-intuitive result. 
When analyzing the currents in Fig.~\ref{fig:fluxes} we find that large values of $g$ (dashed curves) produce smaller currents. 
However, when analyzing the Onsager matrix (which is essentially a measure of the entropy production rate $\Pi$) we see that for certain temperatures the situation is inverted, with larger values of $g$ producing more entropy.

%
%
%
%
\section{\label{sec:disc}Discussions and conclusions}

The goal of this paper was to show how one may construct Lindblad dissipators for a quantum many-body system. 
The main idea is to start with a linear system-bath interaction and then use the method of eigenoperators. 
The resulting dissipator satisfy the two most important property one expects from a system-bath interaction: (i) it correctly takes the system toward the Gibbs thermal equilibrium state and (ii) it does so while satisfying detailed balance. 

We have focused here on the case of a quadratic bosonic chain. 
The reason behind this choice was as follows. 
First, the bosonic nature of the chain makes it more natural to use a  linear system-bath coupling, of the form~(\ref{Hint}). For fermionic systems, on the other hand, difficulties would arise concerning the conservation of particles in the system. 
Second, the quadratic nature of the chain was essential since it makes the problem exactly diagonalizable. 
The main difficulty in the entire procedure is the calculation of the eigenoperators, which require one to know all eigenvalues and eigenvectors of the system Hamiltonian. But for any system that is diagonalizable in a quasi-particle picture, this becomes straightforward. 

All calculations are readily extended to higher dimensions and also to chains involving non-uniform couplings (which may be interesting in the context of disordered systems). 
It should also be possible to extend these results to interacting theories within the mean-field approximation (which essentially amounts to replacing the quartic terms by self-consistent quadratics). 
We therefore view the present work as a first step towards a more systematic approach of constructing Lindblad dissipators for quantum many-body systems. 

\begin{acknowledgements}

G. T. Landi acknowledges the financial support from grant number 2016/08721-7 from the S\~ao Paulo Research Foundation (FAPESP).
J. P. Santos acknowledges the financial support from the CAPES (PNPD program) for the postdoctoral grant.
\end{acknowledgements}

\bibliography{/Users/gtlandi/Documents/library}
\end{document}